\documentclass[11pt]{article}
\usepackage{moriond,epsfig}
\usepackage[tight,scriptsize]{subfigure}
\usepackage[footnotesize]{caption}
\usepackage{floatflt}
\usepackage{wrapfig}

\def\Journal#1#2#3#4{{#1} {\bf #2}, #3 (#4)}

\def\NPB{{\em Nucl. Phys.} B}

\def\PRL{\em Phys. Rev. Lett.}
\def\PRD{{\em Phys. Rev.} D}
\def\PRC{{\em Phys. Rev.} C}

\def\be{\begin{equation}}
\def\ee{\end{equation}}
\def\bea{\begin{eqnarray}}
\def\eea{\end{eqnarray}}

\begin{document}

\vspace*{4cm}
\title{Recent Spin Results from STAR}

\author{Andrew Gordon for the STAR Collaboration}

\address{Department of Physics, Brookhaven National Laboratory,\\
20 Pennsylvania Street, Upton, NY}

\maketitle\abstracts{
In Run 8 at RHIC, STAR significantly enhanced its forward acceptance 
relative to previous years with the 
commissioning of a new detector, the Forward Meson Spectrometer (FMS). 
The large geometrical acceptance of the FMS allows us to extend the forward
reach of the data beyond inclusive pions accessed by modular
calorimeters. The spin-1
$\omega$ is accessible to the FMS through its decay channel
$\omega\rightarrow\pi^0\gamma$. Such events can help disentangle the
dynamical origins of observed large analyzing powers in the forward
region, and can contribute to our knowledge of the nuclear medium
by comparisons of p+p to d+Au. Here we report on the status
of this analysis.}

\section{Introduction}

The fundamental goal of the spin program at STAR is to determine how the
proton acquires spin from its constituent quarks and gluons.
This program makes use of both longitudinally and transversely polarized
p+p collisions at Brookhaven National Laboratory's Relativistic Heavy Ion
Collider (RHIC).
The longitudinal data has allowed STAR to put strong constraints on
the contribution of the gluon spin, down to
x-Bjorken $\sim 0.02$.\cite{long1,long2}

Run 8 at RHIC, which ran during the Fall and Winter of 2007/2008 and
finished in the Spring of 2008, included both d+Au collisions at 
$\sqrt{s_{NN}}=200$ GeV, as well as
transversely polarized p${\uparrow}$ + p${\uparrow}$ collisions at
$\sqrt{s}=200$ GeV. 
Important goals of Run 8 were to provide
measurements of the low-x gluon density in the nucleon, to search for
the onset of gluon saturation effects through intercomparisons of
$\pi^{0}\pi^{0}$ data between p+p and d+Au collisions, and to further
characterize the significant single spin asymmetries that have been
observed in the forward region.\cite{EPJ2005,PRL2008}

Consistent with these goals, STAR commissioned a new detector for Run 8, 
the Forward Meson Spectrometer (FMS).\cite{SPIN08Nikola}
The FMS is a nearly hermetic array of $1264$ lead-glass blocks (``cells'')
situated $\sim 700$ cm downstream of the interaction point and spanning
an area $200\times200$ cm$^2$ perpendicular
to the beam pipe. It covers the full azimuth in
the range $2.5<\eta<4.0$ and provides a many-fold increase in the areal
coverage of the forward region at STAR.\cite{SPIN08Nikola}

STAR has previously reported on precision measurements~\cite{PRL2008}
of the
analyzing power ($A_N$) of inclusive neutral pions in the forward region. 
These measurements used data taken in RHIC runs 3, 5, and 6 with
the Forward Pion Detector (FPD), 
a modular lead-glass array which can be moved horizontally in the
plane transverse to the beamline. These data were taken at
$\sqrt{s}=200$ GeV, where inclusive $\pi^0$ cross sections
are consistent with expectations from pQCD.\cite{PRL2006} The measurements
showed that the variation of $A_N$ with Feynman-x
($X_F=2P_L/\sqrt{s}$) was qualitatively consistent with expectations from
the Sivers effect,\cite{SIVERS1990}
while the $P_T$ dependence was not. Inclusive measurements of
$\eta$ asymmetries have also been reported.\cite{HEPPELMANPANIC08}

The Sivers effect identifies the origin of the observed spin asymmetries with
orbital motion of the quarks inside the polarized proton. This
leads to a correlation between the proton spin and the 
intrinsic transverse momentum of the struck quark in the hard
scatter, which then manifests itself as a Left/Right asymmetry in the resulting
jet direction.  By contrast, in the Collins
effect~\cite{COLLINS1993,CH1994} the 
polarization of the struck quark is correlated to the polarization of the
proton, and the fragmentation of the polarized quark leads to Left/Right
asymmetries within the resulting jet. It remains to be determined the
extent to which these two effects contribute to the observed single spin
asymmetries.\cite{SPIN08Nikola}

Forward jet data can help separate the two contributions. Jet
measurements that integrate the full azimuth
about the jet thrust axis could lead to Sivers-type asymmetries,
while jet measurements that depend 
on the azimuth about the thrust axis could lead to Collins asymmetries.
The Run 8 FMS data were accumulated with a ``high tower'' trigger,
in which a single lead-glass detector was required to have energy above
a threshold to trigger event acquisition. This tends to bias the data
towards jets for which a few electromagnetic particles account for the
bulk of the fragmentation, and PYTHIA\cite{pythia} studies have demonstrated
that we expect most of the energy in these jets to derive from only a few 
fragmentation products. Simulation studies also show resonance
peaks within the observed jet clusters.

The inclusive pions tend to originate from various
resonances along the fragmentation decay chain, and we have
begun to study resonances heavier than neutral pions
as a first step towards understanding jet data.
One source of neutral pions is the isoscaler, spin-1 $\omega$ through
the decay channel $\omega\rightarrow\pi^{0}\gamma$
(BR=8.9\%\cite{PDG}). 
For this resonance, $A_N$ measurements can provide crucial
information for the Collins effect.
The Collins effect is consistent with string
fragmentation models in which a quark/anti-quark pair is produced 
with relative orbital angular momentum at the
point of string breaking.\cite{ARTRU1995,ARTRU1998}
This leads to a Left/Right
asymmetry for the production of the spin 0 pions, and would lead to the
opposite asymmetry for the spin 1 $\omega$.
Direct observation of a negative $A_N$ might provide strong evidence for
the Collins effect.

Finally, there are theoretical expectations that a dense hadronic medium
produces a partial restoration of chiral-symmetry. An observable impact of
this would be shifts in the spectral properties (e.g. mass and width)
of the light vector mesons.\cite{BROWN1992,RAPP2000} An analysis of
data from PHENIX showed no shift in the $\omega$ mass relative to the $\pi^0$, although
the analysis only analyzed central particles and was sensitive down to a minimum
$P_T$ of 2.5 GeV.\cite{ADLER2006}
The d+Au data from Run 8 can potentially allow us to extend this
measurement to higher rapidities and lower $P_T$.

Here we report on initial observations of an $\omega$ signal in the Run 8
p+p data in the FMS.

\section{Data and Simulation}

\begin{wrapfigure}[18]{r}{5.8cm}
\vspace{-.95in}
\hspace{-.19in}
\includegraphics[width=7.5cm]{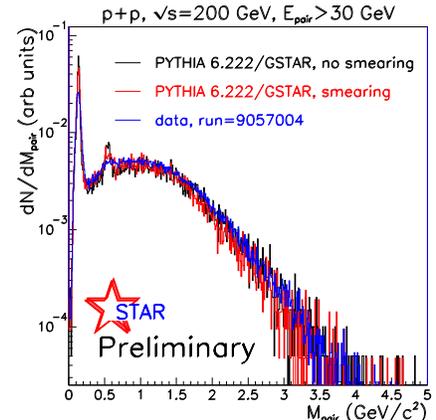}
\noindent \hrulefill
\vspace{-1.2in}
\caption{\label{fig:pairs}
Mass of all pairs with
$E>2$ GeV for p+p data (blue), GEANT simulation (black, labeled ``GSTAR''),
and for GEANT+additional smearing (red).  The
additional smearing is $\sim$ 10 MeV, as
determined from the $\pi^0$ peak in high-tower associated mass distributions.
}
\end{wrapfigure}

The FMS data are clustered, and 
each cluster is fit
to a photon shower shape to determine transverse 
position and energy.
The conversion gains of each cell are determined from
$\pi^0$ mass peaks through an iterative procedure, and these gains have been shown to be
stable over the run at the level of a few percent.
A full PYTHIA(6.222) + GEANT
simulation has been developed,\cite{SPIN08Nikola} and
a comparison of data and simulation is shown in
Figure~\ref{fig:pairs} for the invariant mass of all pairs of clusters
in an event, for minimum bias data. A
pronounced $\pi^0$ mass peak is evident,
and data and simulation agree well over a large range.

Decays of the spin 1, $782$ MeV $\omega$ are accessible to the FMS through the
decay channel $\omega\rightarrow\pi^0\gamma$.
For this decay, two of the photon clusters derive from
the $\pi^0$ decay and one directly from the $\omega$. For each triple of clusters
there are three
possible pairs that can be associated with the $\pi^0$ decay. We form the
invariant mass of each,
and the pair whose
mass is closest to $0.135$ GeV/c$^2$ is associated with
the $\pi^0$ decay. Simulations show that this procedure tags
the photons correctly upwards of $99\%$ of the time. 

To help reduce backgrounds
we apply relatively high thresholds
in both energy and $P_T$. We consider all triples for which
each cluster has energy above 6 GeV,
$|\vec{P}_T(\mathrm{triple})|>2.5$ GeV/c, and $E(\mathrm{triple})>30$ GeV.
We also require that
$P_T>1.5$ GeV/c for the cluster associated with
the $\omega$ decay photon. For the two clusters associated with
the $\omega$ decay pion, we require that their mass be within $0.1$
of $0.135$ GeV/c$^2$ and that $|\vec{P}_T(\pi^0)|>1$ GeV/c.

\begin{wrapfigure}[15]{l}{5.0cm}
\vspace{-1.05in}
\hspace{-.1in}
\includegraphics[width=5.8cm]{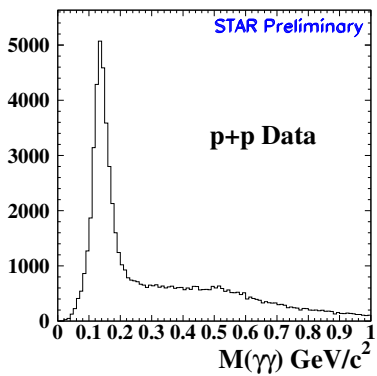}
\vspace{-0.25in}
\caption{
Mass of the two clusters in each triple associated
with the $\omega$ decay $\pi^0$, without $\pi^0$ mass cut.
\label{fig:trippairs}
}
\end{wrapfigure}
Figure~\ref{fig:trippairs} shows the mass of all cluster pairs associated
with the pion in each triple, and a $\pi^0$ peak is evident.

\begin{wrapfigure}[25]{hr}{10.5cm}
\vspace{-2.15 in}
\hspace{-0.1in}
\subfigure
{
 \includegraphics[width=12.0cm]{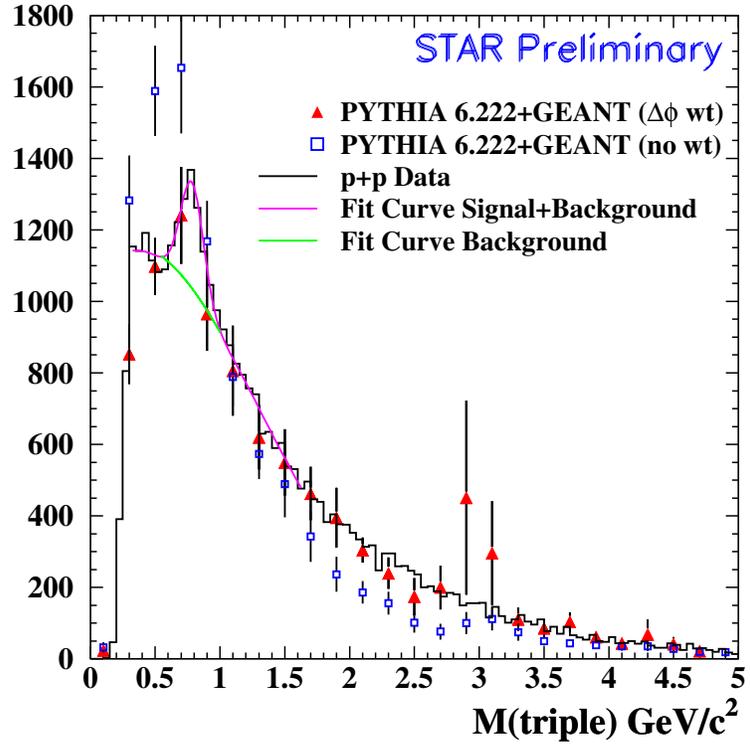}
}\vspace{-0.25in}
\caption{
Mass of all triples for data (black solid),
PYTHIA + GEANT (blue squares), and weighted PYTHIA + GEANT 
(red triangles, see text).
Simulation is normalized to data.
A gaussian+cubic polynomial fit is overlaid (magenta curve).
The fit gaussian mean and width are $\mu=0.784\pm0.008$ GeV/c$^2$ and
$\sigma=0.087\pm0.009$ GeV/c$^2$, and the fitted total area under the Gaussian
is $1339\pm135$ events. The cubic background shape is also shown (green curve).
\label{fig:tripdphimass}
}
\end{wrapfigure}

The mass of all triples is shown in Figure~\ref{fig:tripdphimass}. An
$\omega$ peak of roughly 10 statistical standard deviations is evident.
The PYTHIA(6.222) + GEANT simulation is shown overlaid on the data. The simulation
overpredicts the data at low mass and underpredicts at high mass. Interestingly,
we also observe a discrepancy in the distribution (data not shown) of the quantity
$\Delta\phi\equiv\phi(\pi^0)-\phi(\gamma)$, where $\phi(\pi^0)$ is the azimuth
of the two clusters associated with the $\omega$ decay pion and
$\phi(\gamma)$ the azimuth of the decay photon. Both simulation and data show a single
peak at $0$, but the RMS for the simulation is significantly narrower than the
data ($0.616\pm0.020$ radians for PYTHIA(6.222) + GEANT compared to
$0.827\pm0.003$ for the data, a difference
of $\sim 10$ statistical standard deviations).
To test that $\Delta\phi$
is driving the discrepancy in the mass,
we weighted the simulation
by the ratio of data and
simulation $\Delta\phi$ histograms. The
resulting mass distribution (shown in Figure~\ref{fig:tripdphimass})
agrees well with the data.

To confirm that the $\Delta\phi$ discrepancy is not caused by our detector simulation,
we removed the GEANT simulation
and replaced it with simple geometrical acceptance cuts. We then examined
all PYTHIA photons, treating each as a separate, perfectly measured
electromagnetic cluster. This new $\Delta\phi$ distribution has an
RMS of $0.624\pm0.006$, consistent with the full simulation.
We also examined the PYTHIA parameters of ``CDF
Tune A,''\cite{cdftunea} but saw
no significant change, while changing to PYTHIA version 6.420 produced an
even narrower peak.
We conclude that the distribution of momentum components perpendicular to
the thrust axis ($j_T$) can be tuned in PYTHIA and requires better tuning
in the forward region.

To understand the backgrounds, we examine the
PYTHIA event record for each event in the $\omega$ mass region
($0.68<M(\mathrm{triple})<0.88$ GeV/c$^2$).
The largest background in this region
($\sim 55\pm 10\%$) are events where the three clusters derive from the
decays of two neutral pions. The next
source ($\sim 30\pm 10\%$)
are events that contain both an
$\eta\rightarrow\gamma\gamma$ decay and a $\pi^0$
decay. There are also small contributions from fragmentation
photons as well as other backgrounds.
We note that $\pi^0$
decay photons tend to be near each other at the FMS, and the
$\pi^0\pi^0$

\noindent 
background can be reduced 
by cuts on the smallest transverse distance at the FMS between the
$\omega$ photon and any cluster in the event that is not
part of the triple being analyzed.

\section{Conclusion}

The FMS was newly commissioned for Run 8. The detector has been calibrated,
and a rich data set has begun to be analyzed.
We have reported on an analysis of three-cluster events in the FMS,
with a goal to extract measurements of asymmetries of the spin-1 $\omega$,
as well as
to compare its spectral properties between p+p and d+Au data. 
The $\omega$ signal is readily visible in the p+p data, with a statistical significance of roughly
10 $\sigma$.  However, the signal to noise level is currently insufficient to
extract $A_N$, and future work will focus on attempts to reduce the background level.

\end{document}